\documentclass[amssymb,amsmath,aps,showpacs,twocolumn]{revtex4}
\newcommand{\lsim}{\lesssim}
\newcommand{\gsim}{\gtrsim}

\def\lsim{\mathrel{\raise.3ex\hbox{$<$\kern-.75em\lower1ex\hbox{$\sim$}}}}
\def\gsim{\mathrel{\raise.3ex\hbox{$>$\kern-.75em\lower1ex\hbox{$\sim$}}}}

\def\beq{\begin{equation}}
\def\eeq{\end{equation}}
\def\beqn{\begin{eqnarray}}
\def\eeqn{\end{eqnarray}}
\def\bea{\begin{eqnarray}}
\def\eea{\end{eqnarray}}
\def\be{\begin{equation}}
\def\ee{\end{equation}}
\newcommand{\fslash}[1]{{#1 \kern -0.7em/ \kern 0.1em}}

\begin{document}

\voffset 1.25cm

\title{Spontaneous electro-weak symmetry breaking and cold dark
matter}
\author{ Shou-hua Zhu}
\affiliation{Institute of Theoretical Physics, School of Physics,
Peking University, Beijing 100871, China}

\date{\today}

\begin{abstract}

In the standard model, the weak gauge bosons and fermions obtain
mass after spontaneous electro-weak symmetry breaking, which is
realized through one fundamental scalar field, namely Higgs field.
In this paper we study the simplest scalar cold dark matter model
in which the scalar cold dark matter also obtains mass through
interaction with the weak-doublet Higgs field, the same way as
those of weak gauge bosons and fermions. Our study shows that the
correct cold dark matter relic abundance within $3\sigma$
uncertainty ($ 0.093 < \Omega_{dm} h^2 < 0.129 $) and
experimentally allowed Higgs boson mass ($114.4 \le m_h \le 208$
GeV) constrain the scalar dark matter mass within $48 \le m_S \le
78$ GeV. This result is in excellent agreement with that of W.~de
Boer et.al. ($50 \sim 100$ GeV). Such kind of dark matter
annihilation can account for the observed gamma rays excess
($10\sigma$) at EGRET for energies above 1 GeV in comparison with
the expectations from conventional Galactic models. We also
investigate other phenomenological consequences of this model. For
example, the Higgs boson decays dominantly into scalar cold dark
matter if its mass lies within $48 \sim 64$ GeV.

\end{abstract}


\maketitle

%\section{Introduction}

Understanding the mechanism of electro-weak symmetry breaking (EWSB)
is a primary goal of the Large Hadron Collider (LHC) and
International Linear Collider (ILC). In the standard model (SM),
EWSB is realized through one fundamental Higgs field, namely the
vacuum expectation value (VEV) of Higgs field $<\Phi> \ne 0$. After
EWSB the weak gauge bosons and fermions obtain mass, and only one
neutral Higgs boson is left in particle spectrum. The couplings
among Higgs boson and weak gauge bosons/fermions are proportional to
their corresponding mass. This is one of the most important features
of spontaneous EWSB. Though this feature has not been directly
tested in the past experiments, the SM predictions and experimental
measurements are in good agrement at an accuracy of $O(0.1\%)$. The
global fit of LEP, SLD and Tevatron data predicts the Higgs boson
mass to be $m_h=98^{+52}_{-36}$ GeV and $m_h <208$ GeV at 95\% CL
using latest preliminary top quark mass $m_t=174.3\pm 3.4$ GeV
\cite{Juste:2005dg}. The latest direct search sets the lower bound
of SM Higgs boson of $114.4$ GeV at 95\% confidence level (CL)
\cite{Barate:2003sz}. Because EWSB is the only untouched part in the
SM, it is most likely that the new physics is closely related to the
Higgs sector.

Over the past several years our understanding of components of the
Universe has undergone important advances. The existence of dark
matter (DM) is by now well established. The non-baryonic cold dark
matter density is (within $3\sigma$ uncertainty)
\cite{Eidelman:2004wy}
\begin{eqnarray}
0.093 < \Omega_{dm} h^2 < 0.129 \label{dmdensity}
\end{eqnarray}
where $h \approx 0.71$ is the normalized Hubble expansion rate.
However the microscopic properties of dark matter are remarkably
unconstrained. Therefore it is quite interesting to ask how cold
dark matter interacts with usual SM matter. Though the cold dark
matter is allowed to feel weak interaction, such possibility will
usually involve more free parameters. The popular cold dark matter
candidate neutralino in supersymmetrical models is a good example.
Therefore in this paper we only discuss the case that dark matter is
SM gauge group singlet.

Motivated by the successful assumption that weak gauge bosons and
fermions obtain mass from the interactions with weak-doublet Higgs
field, it is quite natural to assume that SM gauge group singlet
cold dark matter obtains mass the same way as those of weak gauge
bosons and fermions. Thus the coupling among Higgs boson and cold
dark matter is solely fixed by dark matter mass.  If the dark matter
is fermion, it will obtain mass through dimension-five
non-renormalizeble operator $(\overline{\Psi} \Psi) (\Phi^+\Phi)$
with $\Phi$ the usual weak-doublet Higgs field. Thus we prefer to
take the cold dark matter as scalar field, denoted as S. In fact the
scalar particle as cold dark matter has been widely studied in
literature \cite{scalarcdm,Burgess:2000yq}. If we introduce only one
singlet scalar field S, there is only one extra free parameter other
than those in the SM, namely the mass of the scalar cold dark matter
$m_S$. In this paper we will study this simplest cold scalar dark
matter model. Our results show that the correct cold dark matter
relic density in Eq. (\ref{dmdensity}) and experimentally allowed
Higgs boson mass ($114.4 \le m_h \le 208$ GeV \footnote{Because the
singlet scalar only interacts with Higgs boson, current electroweak
precision data is not sensitive to such kind of new interactions
beyond the SM. Therefore we assume that experimentally allowed Higgs
boson mass in the SM is not affected.} ) constrain the scalar dark
matter within $48 \le m_S \le 78$ GeV. This result is in excellent
agreement with that of W.~de Boer et.al. ($50 \sim 100$ GeV)
\cite{deBoer:2005tm}. Such kind of dark matter annihilation can
account for the observed gamma rays excess ($10\sigma$) at EGRET for
energies above 1 GeV in comparison with the expectations from
conventional Galactic models. In this paper we also investigate
other phenomenological consequences of this model. For example, the
Higgs boson decays dominantly into scalar cold dark matter if its
mass lies within $48 \sim 64$ GeV.

%\section{Model}

The Lagrangian of the simplest model can be written as
\begin{eqnarray}
L=L_{SM}+ \frac{1}{2} \partial_\mu S \partial^\mu S
-\frac{\lambda_S}{4} S^4-\lambda S^2 (\Phi^+ \Phi)
\label{model}
\end{eqnarray}
where $L_{SM}$ is the Lagrangian of the SM and $\Phi$ is the weak
doublet Higgs field. $L$ is obviously invariant under discrete
transformation $S \rightarrow -S$, which ensures S the good
candidate of cold dark matter. The model discussed in this paper
is different from that of Ref. \cite{Burgess:2000yq}. The main
difference is that the mass term of S field $m_0^2 S^2$ is not
included in this paper. It is known that the mass of dark matter
can vary many orders of magnitude depending on the strength it
interacts with usual matter. Here we investigate the possibility
that dark matter is in the weak scale while $m_0^2$ can be
anything. Therefore it is unnecessary to introduce one extra new
mass scale. In this paper we assume that weak doublet Higgs field
{\em solely} induces mass for all particles: weak gauge bosons,
fermions and singlet scalar field $S$. The model can be a massless
one before EWSB. The spontaneous symmetry breaking, trigger by
$\mu^2 \Phi^+\Phi$, can be induced radiatively
\cite{Coleman:1973jx}. On the contrary $S^2$ term is absent or
negligibly small provided that it does not get contribution from
$\lambda S^2 (\Phi^+ \Phi)$.
 After spontaneous electro-weak symmetry
 breaking $<\Phi>=v=246$ GeV, the Higgs boson, as in
the standard model, $m^2_h=\lambda_h v^2$ with $ \lambda_h$ the
coefficient of $(\Phi^+\Phi)^2$ and $m^2_S=\lambda v^2$. It is
obvious that coupling $\lambda$ is determined by $m_S$ and in this
model  $\lambda$ is the only extra free parameter relevant to our
discussion.

%\section{Relic density calculation}

Next we will explore the cosmological constraints on this model by
demanding the present abundance of S to be in the range of Eq.
(\ref{dmdensity}). As we can see, this imposes a very strong
relationship between Higgs boson mass and $m_S$. The calculation of
S abundance follows the standard procedure \cite{KT} and we refer
the interesting reader to Ref. \cite{Burgess:2000yq} for the
details.

The present density of S can be written as \cite{Burgess:2000yq}
\begin{eqnarray}
\Omega_S h^2 = \frac{(1.07 \times 10^9) x_f}{\sqrt{g_*} M_{pl}[{\rm
in \ GeV}] <\sigma v_{rel}>}, \label{density}
\end{eqnarray}
where $g_*$ counts the degrees of freedom in equilibrium at
annihilation, $x_f$ is the inverse freeze-out temperature in units of
$m_S$, which can be obtained by solving the equation
\begin{eqnarray}
x_f\simeq \ln\left[\frac{ 0.038  M_{pl} m_S <\sigma
v_{rel}>}{\sqrt{g_* x_f}}\right]. \label{xf}
\end{eqnarray}
In Eqs. (\ref{density}) and (\ref{xf}), $v_{rel}$ is the relative
velocity of the two incoming dark matter particles, $M_{pl}$ is the
Planck mass and $<...>$ denotes the relevant thermal average.

\begin{figure}[thb]
\vbox{\kern2.0in\includegraphics{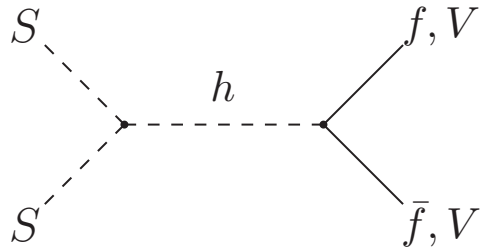}} \caption{
 Schematic Feynman diagram for $SS \rightarrow $ SM
particles. Here f and V represent SM fermions and weak gauge bosons
respectively. } \label{fig1}
\end{figure}

Since the scalar dark matter obtains mass through interaction with
VEV of the Higgs field, it is natural to expect that dark matter
mass scale is $O(100)$ GeV. Therefore $\sigma v_{rel}$ can be
obtained by evaluating the tree-level diagram in Fig. \ref{fig1}. In
the non-relativistic limit, $\sigma v_{rel}$ is
\cite{Burgess:2000yq}
\begin{eqnarray}
\sigma_{ann} v_{rel} &=& \frac{8 \lambda^2 v^2}{(4
m_S^2-m_h^2)^2+m_h^2 \Gamma_h^2}\ F_X \nonumber \\
&=& \frac{8 m_S^4 }{v^2 \left[(4 m_S^2-m_h^2)^2+m_h^2
\Gamma_h^2\right]}\ F_X \label{sigamv}
\end{eqnarray}
with
\begin{eqnarray}
F_X = \lim_{ m_{\tilde{h}} \rightarrow 2 m_S } \left(
\frac{\Gamma_{\tilde{h}\rightarrow X} }{m_{\tilde{h}}} \right).
\end{eqnarray}
Here $\Gamma_h$ is the Higgs total decay width and
$\Gamma_{\tilde{h}\rightarrow X}$ denotes the partial decay width
for the virtual $\tilde{h}$ decay into $X$, $\tilde{h} \rightarrow
X$, in the limit $m_{\tilde{h}} \rightarrow 2 m_S$. Here X
represents SM particles.

%\section{Constraints on dark matter mass and discussions}

\begin{figure}[thb]
\vbox{\kern3.0in\includegraphics{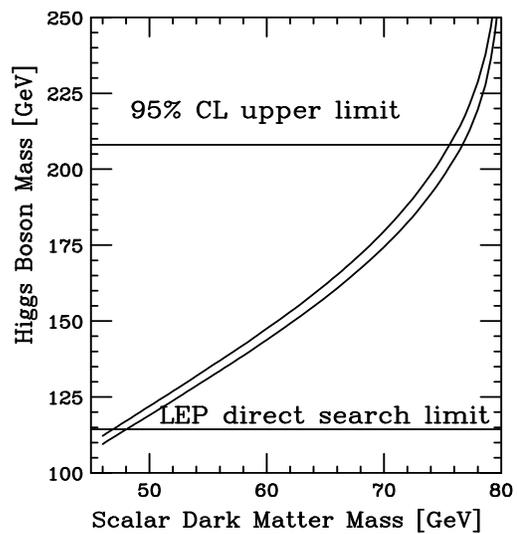}} \caption{ Allowed Higgs boson
and scalar dark matter mass region (between two curves). This mass
region can produce correct relic density $ 0.093 < \Omega_{dm} h^2
< 0.129 $ (within $3\sigma$ uncertainty). Also shown are Higgs
boson upper (208 GeV) and lower (114.4 GeV) mass limits from
electro-weak precision data global fit and direct search at LEP.}
\label{fig2}
\end{figure}

In Fig. \ref{fig2} and table \ref{table1} we show the allowed Higgs
boson and scalar dark matter mass region [between two curves],
provided that relic abundance satisfies Eq. (\ref{dmdensity}). From
figure we can see a very strong relationship between Higgs boson
mass and $m_S$. It is known that the coupling strength among Higgs
boson and SM particles is proportional to their corresponding mass,
as we discussed above. At the same time, the coupling between Higgs
boson and scalar dark matter is also proportional to $m_S$. Once
$m_S$ is fixed, the only free varying parameter is $m_h$. If we
impose the Higgs mass constraints from direct search limit 114.4 GeV
to 208 GeV, the allowed scalar dark matter is
\begin{eqnarray}
48 \le m_S \le 78 \ GeV. \label{massregion}
\end{eqnarray}
It should be noted that $m_S$ can't be larger than  $m_W$. Otherwise
the cross section of $SS \rightarrow WW$ becomes too large and the
relic abundance is out of the region in Eq. (\ref{dmdensity}). Eq.
(\ref{massregion}) is in excellent agreement with that of W.~de Boer
et.al. ($50 \sim 100$ GeV) \cite{deBoer:2005tm}. Such kind of dark
matter annihilation produces mono-energetic quarks of $50 \sim 100$
GeV, which can account for the observed gamma rays excess
($10\sigma$) at EGRET for energies above 1 GeV in comparison with
the expectations from conventional Galactic models. It should be
emphasized that the crucial part of EGRET photon excess origin is
mono-energetic quarks, no matter which kind of dark matter
annihilation produces them. In fact the further investigation by
W.~de Boer et.al. \cite{deBoer:2005bd} shows that supersymmetric
models with suitable parameter space are compatible with their
conjecture. There the dark matter candidate is neutralino.

It is also interesting to note that for Eq. (\ref{massregion}),
$F_{VV^*}$ with $V^*$ the virtual $W$ and $Z$ plays a very important
role. $F_{VV^*}$ can be extracted from Ref. \cite{Djouadi:2005gi} as
\begin{eqnarray}
F_{VV^*}= \frac{3 G_F^2 m_V^4}{16 \pi^3} \delta _V R(x)
\end{eqnarray}
with $\delta_W=1, \delta_Z=7/12-10/9 \sin^2\theta_W+40/9
\sin^4\theta_W$ ($\theta_W$ is the weak angle). Here
\begin{eqnarray}
R(x)&=&\frac{3 (1-8 x+20 x^2)}{\sqrt{4 x-1}} \arccos\left(\frac{3
x-1}{2 x^{3/2}} \right) \nonumber \\
&& -\frac{1-x}{2 x} (2- 13 x+ 47 x^2) \nonumber
\\
&& -\frac{3}{2} (1- 6 x+ 4 x^2) \log{x}
\end{eqnarray}
with $x=m_V^2/(4 m^2_S)$.

\begin{table}[htb]
\begin{tabular}{l|cc}
\hline \hline
$m_S$  [GeV] & $m_h$ upper limit [GeV] &  $m_h$ lower limit [GeV] \\
\hline
50 & 122 & 119 \\
55 & 134 & 131 \\
60  & 148 & 144  \\
65  & 162 & 158  \\
70 & 180 & 174 \\
75 & 204 & 197 \\
80 & 275 & 261 \\
 \hline \hline
\end{tabular}
\caption{ Upper and lower limits on Higgs boson for several $m_S$ in
order to obtain the correct relic abundance in Eq. (\ref{dmdensity}).
} \label{table1}
\end{table}

\begin{figure}[thb]
\vbox{\kern3.0in\includegraphics{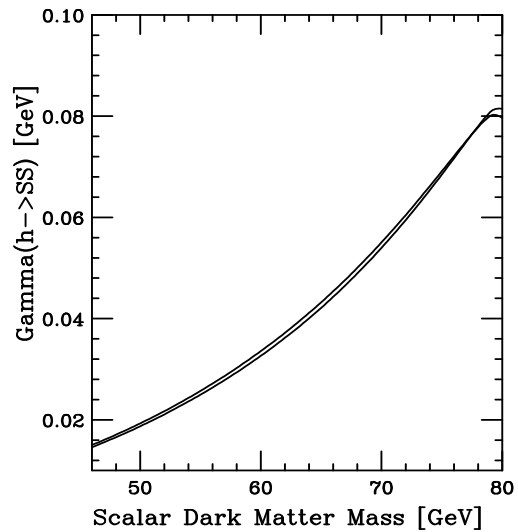}} \caption{ Decay width [in GeV] of $h
\rightarrow SS$ as a function of $m_S$.
Region between two curves can produce correct
cold dark matter relic abundance in Eq.(\ref{dmdensity}).
} \label{fig3}
\end{figure}

\begin{figure}[thb]
\vbox{\kern3.0in\includegraphics{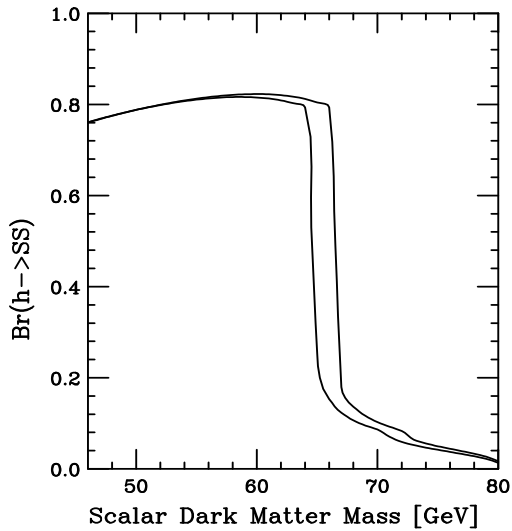}} \caption{ Branching ratio of $h \rightarrow
SS$ as a function of $m_S$. Other conventions are the same with
Fig. \ref{fig3}. } \label{fig4}
\end{figure}

Fig.~\ref{fig2} and table \ref{table1} also show that in the allowed
mass region, $m_h
> 2 m_S$. This feature has very important phenomenological consequence, namely the
Higgs can decay into pair of dark matter. The partial decay width
can be written as
\begin{eqnarray}
\Gamma(h\rightarrow SS)=\frac{m_S^4}{8 \pi v^2 m_h} \sqrt{1-\frac{4
m_S^2}{m_h^2}}.
\end{eqnarray}
$\Gamma(h\rightarrow SS)$  and $Br(h \rightarrow SS)$ are shown in
Fig. \ref{fig3} and \ref{fig4}. From Fig. \ref{fig3} we can see that
the partial decay width is between $0.02 \sim 0.08$ GeV for $m_S= 48
\sim 78$ GeV. In Fig. \ref{fig4} the rapid drop around $m_S=65$ GeV
is due to the rapid growth of $\Gamma(H\rightarrow VV^{*})$. From
Fig. \ref{fig4} we can see good and bad news. The good news is that
there is certain possibility that Higgs boson can decay into SM
particle, especially for $m_S \ge 64 $ GeV. Therefore we still have
opportunity to see Higgs boson resonance as the case in the SM. The
bad news is that we need more luminosity for the usual Higgs search
strategies, especially for $m_S \le 60$ GeV. Further efforts should
be put on search of the invisibly Higgs boson decay
\cite{invisible}.

In this paper we study the simplest scalar cold dark matter model
in which the scalar cold dark matter obtains mass through
interaction with the weak-doublet Higgs field, the same way as
those of weak gauge bosons and fermions. The coupling between
Higgs boson and scalar dark matter is solely fixed by scalar dark
matter mass. The correct cold dark matter relic abundance  within
$3\sigma$ uncertainty ($ 0.093 < \Omega_{dm} h^2 < 0.129 $) and
experimentally allowed Higgs boson mass ($114.4 \le m_h \le 208$
GeV) constrain the scalar dark matter mass within $48 \le m_S \le
78$ GeV. This result is in excellent agreement with that of W.~de
Boer et.al. ($50 \sim 100$ GeV). Such kind of dark matter
annihilation can account for the observed gamma rays excess
($10\sigma$) at EGRET for energies above 1 GeV in comparison with
the expectations from conventional Galactic models. We also
investigate other phenomenological consequences of this model.
Most importantly the Higgs boson decays dominantly into scalar
cold dark matter if its mass lies within $48 \sim 64$ GeV.

Recently Bergstrom et. al. \cite{Bergstrom:2006tk} investigated
the internal consistency of the halo dark matter model which has
been proposed by de Boer et. al. Certain inconsistency is found
\cite{Bergstrom:2006tk}. Now that the proposal of de Boer et. al.
is still in debate, we should emphasize that our main results,
i.e. (1) the correlation between masses of scalar dark matter and
Higgs boson, (2) the Higgs boson can decay into dark matter with
large branching ratio, might be irrelevant to the EGRET data.
However LHC/ILC can provide ideal place to investigate the scalar
dark matter in the Higgs boson decay \cite{invisible}.

\noindent {\em Acknowledgements}: The author thanks Prof. Y.Q. Ma for
the discussion on EGRET, and Prof. C. Liu and H.Q. Zheng for the
discussion on Eq. (\ref{model}). This work was
supported in part by the Natural Sciences Foundation of China under
grant No. 90403004, the trans-century fund and
the key grant project (under No. 305001) of Chinese Ministry of
Education.


\begin{thebibliography}{99}

%\cite{Juste:2005dg}
\bibitem{Juste:2005dg}
  A.~Juste,
  %``Top quark measurements,''
  arXiv:hep-ex/0511025.
  %%CITATION = HEP-EX 0511025;%%

\bibitem{Barate:2003sz}
  R.~Barate {\it et al.}  [LEP Working Group for Higgs boson searches],
  %``Search for the standard model Higgs boson at LEP,''
  Phys.\ Lett.\ B {\bf 565}, 61 (2003).
%  [arXiv:hep-ex/0306033].
  %%CITATION = HEP-EX 0306033;%%

%\cite{Eidelman:2004wy}
\bibitem{Eidelman:2004wy}
  S.~Eidelman {\it et al.}  [Particle Data Group],
  %``Review of particle physics,''
  Phys.\ Lett.\ B {\bf 592} (2004) 1.
  %%CITATION = PHLTA,B592,1;%%


\bibitem{scalarcdm}
 M.~J.~G.~Veltman and F.~J.~Yndurain,
  %``Radiative Corrections To W W Scattering,''
  Nucl.\ Phys.\ B {\bf 325}, 1 (1989);
  %%CITATION = NUPHA,B325,1;%%
  %%Cited 62 times in SPIRES-HEP
V.~Silveira and A.~Zee,
  %``Scalar Phantoms,''
  Phys.\ Lett.\ B {\bf 161}, 136 (1985);
  %%CITATION = PHLTA,B161,136;%%
  %%Cited 14 times in SPIRES-HEP
  J.~McDonald,
  %``Gauge singlet scalars as cold dark matter,''
  Phys.\ Rev.\ D {\bf 50}, 3637 (1994).
  %%CITATION = PHRVA,D50,3637;%%
  %%Cited 16 times in SPIRES-HEP
  H.~Davoudiasl, R.~Kitano, T.~Li and H.~Murayama,
 %  ``The new minimal standard model,''
  %
  Phys.\ Lett.\ B {\bf 609}, 117 (2005).
%  [arXiv:hep-ph/0405097].
  %%CITATION = HEP-PH 0405097;%%

%\cite{Burgess:2000yq}
\bibitem{Burgess:2000yq}
  C.~P.~Burgess, M.~Pospelov and T.~ter Veldhuis,
  %``The minimal model of nonbaryonic dark matter: A singlet scalar,''
  Nucl.\ Phys.\ B {\bf 619}, 709 (2001)
  [arXiv:hep-ph/0011335].
  %%CITATION = HEP-PH 0011335;%%
  %%Cited 21 time in SPIRES-HEP

%\cite{deBoer:2005tm}
\bibitem{deBoer:2005tm}
  W.~de Boer, C.~Sander, A.~V.~Gladyshev and D.~I.~Kazakov,
  %``EGRET excess of diffuse galactic gamma rays as tracer of dark matter,''
  arXiv:astro-ph/0508617.
  %%CITATION = ASTRO-PH 0508617;%%

%\cite{Coleman:1973jx}
\bibitem{Coleman:1973jx}
  S.~R.~Coleman and E.~Weinberg,
  %``Radiative Corrections As The Origin Of Spontaneous Symmetry Breaking,''
  Phys.\ Rev.\ D {\bf 7}, 1888 (1973).
  %%CITATION = PHRVA,D7,1888;%%
  %%Cited 745 times in SPIRES-HEP


\bibitem{KT}
E.W. Kolb and M.S. Turner, The Early Universe, Addison-Wesley, 1990.

%\cite{deBoer:2005bd}
\bibitem{deBoer:2005bd}
  W.~de Boer, C.~Sander, V.~Zhukov, A.~V.~Gladyshev and D.~I.~Kazakov,
  %``The supersymmetric interpretation of the EGRET excess of diffuse galactic
  %gamma rays,''
  arXiv:hep-ph/0511154.
  %%CITATION = HEP-PH 0511154;%%
%\cite{Djouadi:2005gi}
\bibitem{Djouadi:2005gi}
  A.~Djouadi,
  %``The anatomy of electro-weak symmetry breaking. I: The Higgs boson in  the
  %standard model,''
  arXiv:hep-ph/0503172 and references therein.
  %%CITATION = HEP-PH 0503172;%%
  %%Cited 26 times in SPIRES-HEP

\bibitem{invisible}
 S.~h.~Zhu,
  %``Detecting an invisible Higgs boson at Fermilab Tevatron and CERN LHC,''
  Eur.\ Phys.\ J.\ C {\bf 47}, 833 (2006)
  [arXiv:hep-ph/0512055]
  %%CITATION = HEP-PH 0512055;%%
 and references therein.

%\cite{Bergstrom:2006tk}
\bibitem{Bergstrom:2006tk}
  L.~Bergstrom, J.~Edsjo, M.~Gustafsson and P.~Salati,
 %  ``Is the dark matter interpretation of the EGRET gamma excess compatible with
  %antiproton measurements?,''
  JCAP {\bf 0605}, 006 (2006)
  [arXiv:astro-ph/0602632]; for the earilier invesigations to see,
  for exmple, A.~Cesarini, F.~Fucito, A.~Lionetto, A.~Morselli and P.~Ullio,
  %``The galactic center as a dark matter gamma-ray source,''
  Astropart.\ Phys.\  {\bf 21}, 267 (2004)
  [arXiv:astro-ph/0305075]; L.~Bergstrom, P.~Ullio and J.~H.~Buckley,
  % ``Observability of gamma rays from dark matter neutralino annihilations  in
  %the Milky Way halo,''
  Astropart.\ Phys.\  {\bf 9}, 137 (1998)
  [arXiv:astro-ph/9712318];
  %%CITATION = ASTRO-PH 0602632;%%
A.~Bottino, F.~Donato, N.~Fornengo and S.~Scopel,
  % ``Indirect signals from light neutralinos in supersymmetric models  without
  %gaugino mass unification,''
  Phys.\ Rev.\ D {\bf 70}, 015005 (2004)
  [arXiv:hep-ph/0401186].



\end{thebibliography}
\end{document}